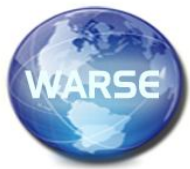

# Development of a Secure and Private Electronic Procurement System based on Blockchain Implementation

August Thio-ac[1], Erwin John Domingo[2], Ricca May Reyes[3], Nilo Arago[4], Romeo Jr. Jorda[5], Jessica Velasco[6]
[1,2,3,4,5,6]Electronics Engineering Department, Technological University of the Philippines, Philippines

**ABSTRACT**

This paper presents the development of an online procurement system and the integration of blockchain technology. Various tools such as PHP, JavaScript, HTML, CSS, and jQuery were used in designing the graphical, programming logic, and blockchain aspect of the system. Every page and function will have their respective construction and result. In addition, the proposed system's flow of process and the methods on the testing and hosting of the site as well as the different web development languages used in every part of the development and design process were presented. The proposed system was successfully and functionally developed starting from the execution of procurement proper, to the placement of procured items or goods, and up to the signing of contracts by the winner and the procurer. Lastly, features were added such as user profiles of the bidder and procurer.

## 1. INTRODUCTION

The integration of blockchain results to different benefits in every aspect of any type of digital system primarily in the side of security while also improving the business side [1]. Blockchain can be used in an electronic voting system for security and account purposes. Blockchain integration has many from planning and designing, to creating nodes for different institutions to distribute [2]-[6]. Moreover, blockchain technology can be applied to different or various fields of business. This paper explains on how blockchain technology works when people get their health documents as supposed to paper system. This allows us to establish the benefits of using blockchain technology in an electronic procurement system [7].

The process and the technical aspects of procurement, from purchase up to declaration of winners, are represented through various mathematical equations with different situations and various ways on how to procure items and goods [8]. However, electronic procurement systems cannot be perfect, there will always be limitations on what function it can perform. The traditional paper system will always have advantages over electronic procurement [9]. Blockchain integration may help, but it will always have limitations.

The development of an online system can be compared to the construction of a building. Before the construction, an architect develops a design of the interior and exterior of the building. Then an engineer or designer will start with the construction and comply with the legal requirements for the development thereof. Development, design, documentation, content creation, programming, blockchain integration, and online deployment requires specific tools and techniques for a platform to be constructed according to the plan [10].

If blockchain will be integrated with the Internet of Things (IoT), faster in speed, higher security, and easier traceability will be expected in the rest of the supply chain. Internet of Things can interrelate with the accessible input data resources and probable output applications whether in medical, energy, or agricultural automation applications together through the Internet [11]-[16].

## 2. RESEARCH METHODOLOGY

The methodology used in different aspects of the system requires different way of thinking, planning, and approach respectively. Different functions are constructed using different languages, tools, and techniques. Different ideas should be implemented in their own way [17].

### 2.1 Problem Definition

The main problems encountered in the development of the system are focused on the systemization of the procurement process as supposed to the traditional private procedures and the limitations of the functions that will be discussed on the next sections of this paper.

### 2.2 Requirement Assessment

To be able to build an online electronic procurement system as shown by the process flowchart on Figure 1, the functions were the most crucial part. PHP and JavaScript were needed to build the programming logic and design of the functions and blockchain integration for the procurement processes while HTML, CSS, and jQuery were used for the aesthetics and graphical design.

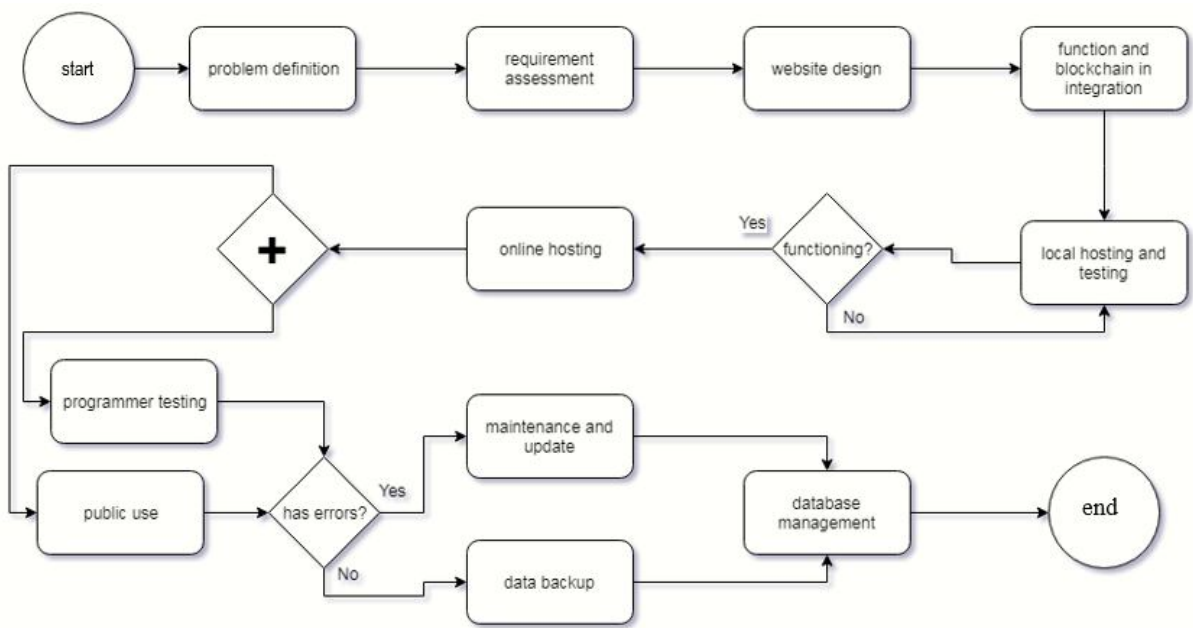

**Figure 1:** Process flowchart

### 2.3 Website Design, Functions, and Blockchain Integration

The complexity of a procurement system as shown in Figure 2 requires a carefully thought out methodology. The design was implemented at the start for the ease of editing while integrating the blockchain and the programming logic and design.

Simple layouts of web pages were created with HTML and CSS while using jQuery for triggers and animations. XAMPP as seen in Figure 3 was used for building the system locally and testing the functionality of different programming logic. The index page was built with aesthetics as the priority; this was to please the users while using the site for the first time. The account system was the first aspect that was built with programming logic and design; the users should be able to log in and sign up at the index page to be convenient and straightforward. A different page for the users' profiles was built. The blockchain integration connected with the account system will be the sending of XEM tokens, which the users, use for bidding purposes.

Continuing the account system, users' profiles can be edited with their input information and a dedicated page for editing their respective profiles. A button for uploading their Know Your Customer (Valid ID's for the majority while students can just use their school ID) can be seen clearly for the users to realize the importance of it.

The KYC will allow other users and the administrators as well, to identify if the user's identity is legit and verify them to the system [15], allowing them to interact with others and join the procurement processes, or maybe start one. Signing up, an email using SMTP is sent to the users' emails to verify their account and allowing them to download their private key in the instructions page. This private key is in JSON file format and is used in bidding and signing contracts.

The market page, or shop, is where the procurement processes lie on. The design of the page was like a simple e-commerce site that is convenient and straightforward. The design was built with more CSS with the help of Bootstrap and jQuery.

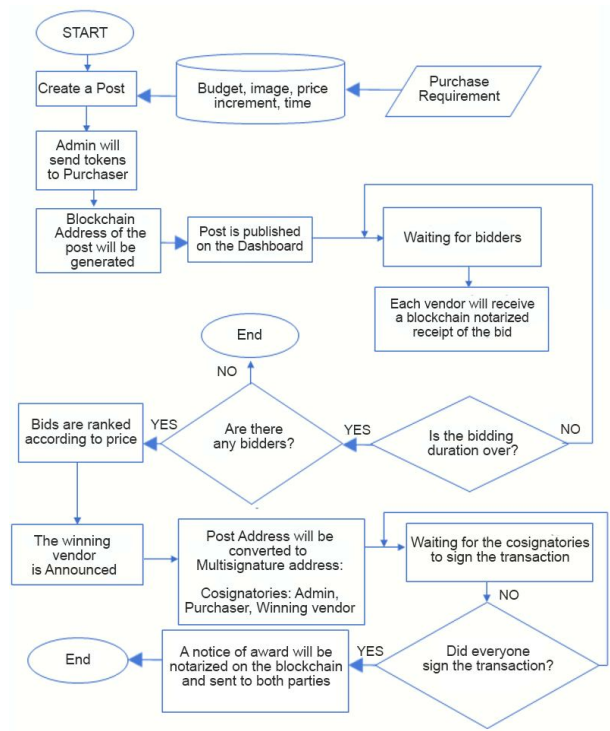

**Figure 2:** Procurement process

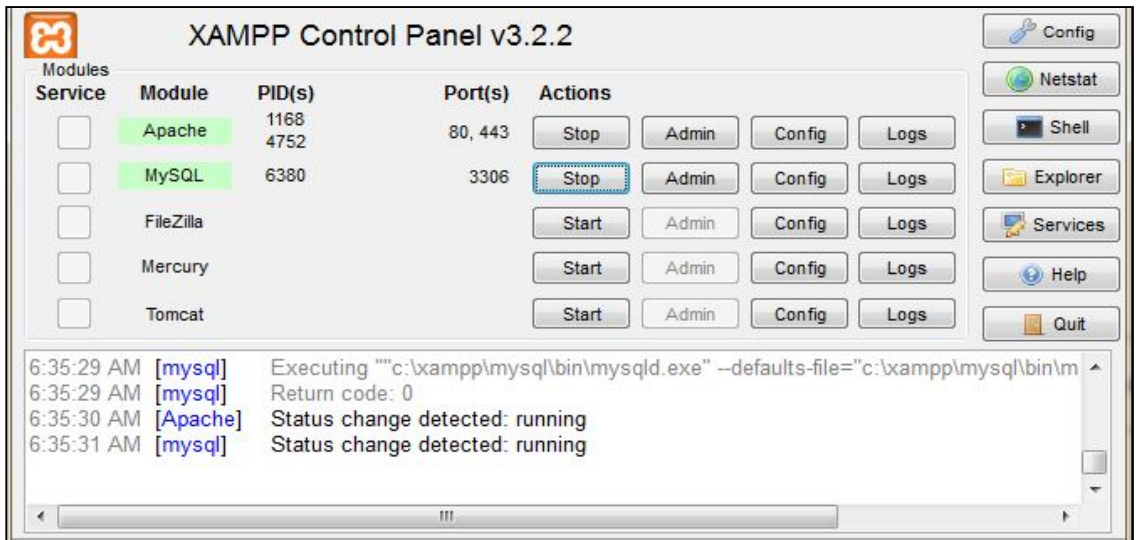

**Figure 3:** XAMPP Interface

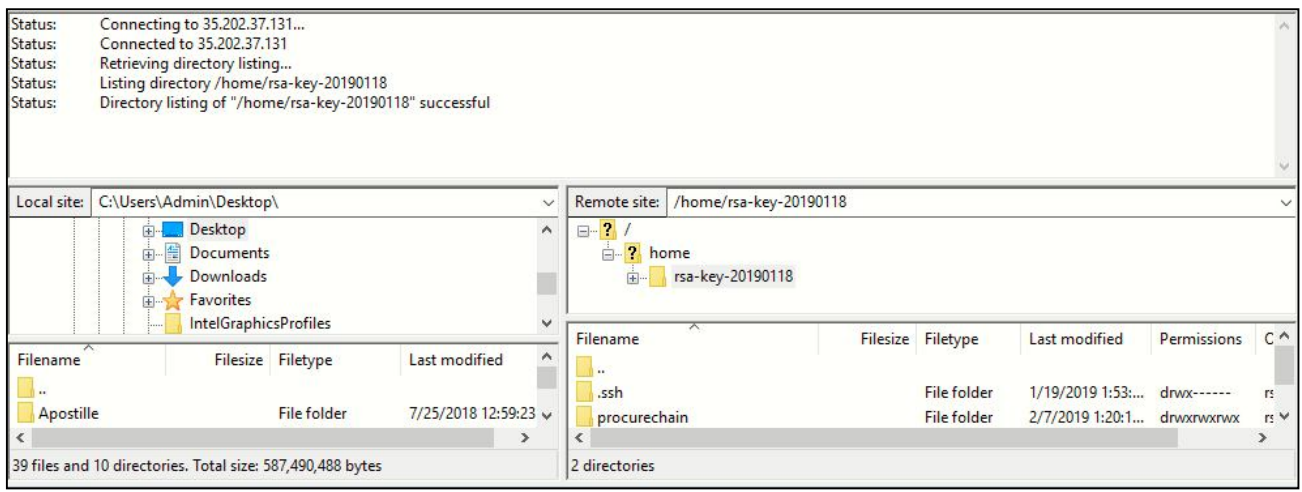

**Figure 4:** FileZilla FTP client Interface

The programming logic and design used in this page was the declaration of procurement winners and the listing of procurement items or services. The integration of blockchain in this page was the most crucial part. Putting the scripts that create the multi-signature wallets and sending of contracts to the users and posts were created using JavaScript and PHP. The creation posts can also be seen and used in this page, which is the most basic part of the procurement processes that occur in the system.

**2.4 Local and Online Hosting**

The creation of the system started with designing the index and market page using HTML, CSS, and jQuery. The functions were written in PHP, while the blockchain functions were written in JavaScript. XAMPP is a tool used for local hosting the system, it lets the device running it be a local server and other users can connect through WIFI or WLAN. XAMPP is a free open source software that allows devices to interpret programs that were written in PHP. This method allows the programmers to test the system locally and detect any bugs and loopholes in it.

**2.5 Programmer Testing and Public Use**

An online system cannot be perfect at the first period it was hosted. Bugs and errors will always prevail even after the programmers repeatedly tested the whole site and its functions. Users can communicate with the programmers personally or through the site's email address so they can troubleshoot everything and update the files.

The different components of the site were link after testing it locally, and online using local and cloud servers. Even after completing every file and page of the site, proper testing was considered since bugs and errors will always occur. Maintenance and update is still a priority for the programmers.

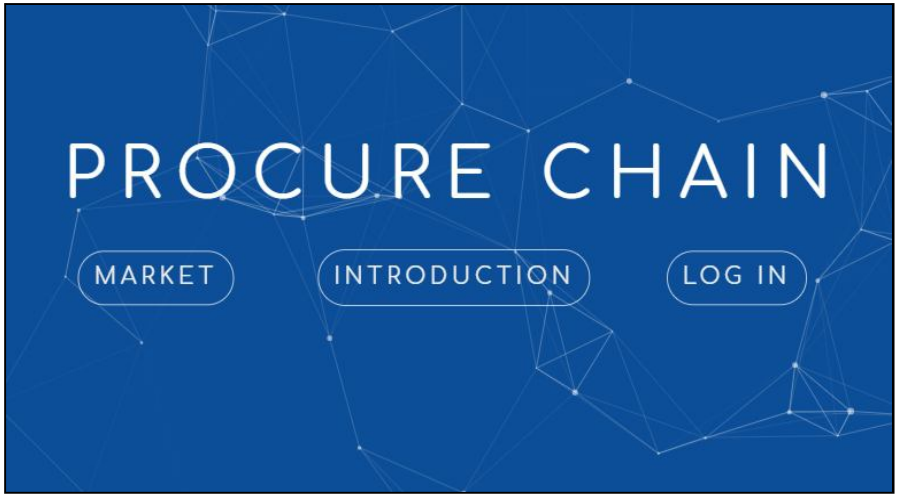
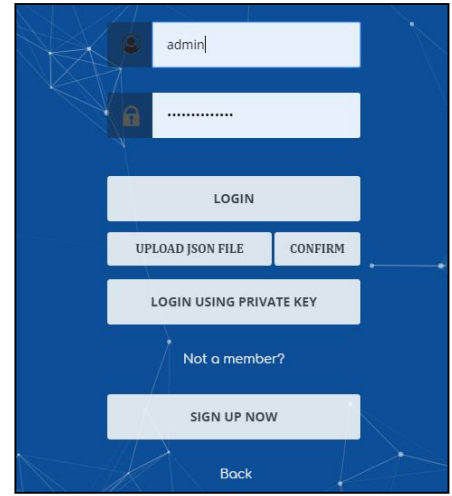

**Figure 5:** Procure Chain Index page  **Figure 6:** Procure Chain Login

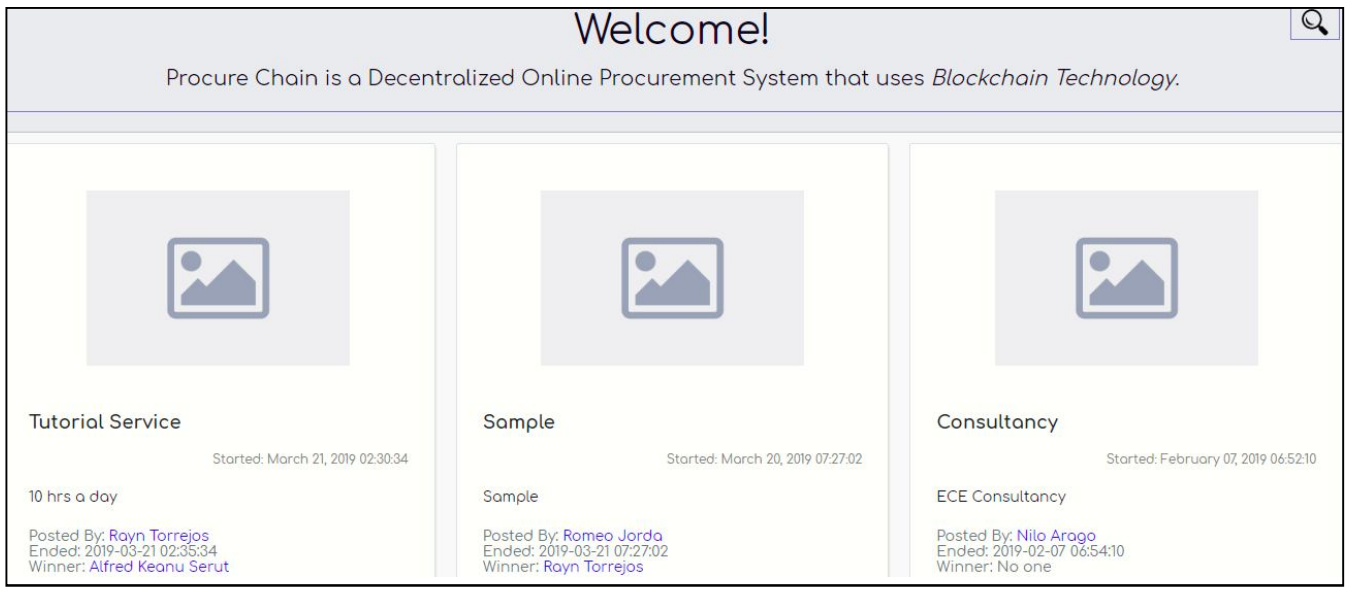

**Figure 7:** Procure Chain Market page

**2.6 Maintenance and Update**

Every user can experience some peculiar events on our site. Some devices are specifically incompatible with the scripts and functions or the aesthetics of the site. Changes and upgrades can always occur; this is where maintenance and update take place with the old files being changed and new files being added for the compatibility and functionality of the site. Files can always be changed or added easily through the FTP client and server software, FileZilla (Figure 4).

**3.1 Data Backup and Database Management**

Unwanted errors and events can and will always occur. User mistakes happen often, and data will always be at risk, whether incorrect data input or details, backups will always be important for the safety of other data. Database management is a priority for the security and safety of user information, as well as the subjects that are involved with the processes of the functions of the site. Data backups and checking are executed from time to time; typically, every day.

**3. RESULT**

A blockchain based procurement platform was developed using three blockchain components – digital signature, multisignature protocol, and blockchain notarization. The blockchain platform only focuses on blockchain integration in a procurement process. The program is limited to the ranking of supplier's bid according to price. Figure 5 shows the index page while Figure 6 shows the login page of the developed Procure Chain (blockchain-based procurement platform).

Several private entities were asked to use the platform, focusing on its key blockchain features. An evaluation sheet is used to collect the response of the users based on their experience using the platform. In addition to the collection of data from private individuals, an interview with four government procurement related entities were conducted.

**3.2 Market & Profile Page**

The same programming languages were used to build the market page. The market page (Figure 7) consists of the shop window, which shows every items or services being procured, the create post button that allows verified users to procure, and the blockchain scripts for the posts that already ended.

The profile and edit profile page (Figure 8) heavily depend on the database of the users. Database management was used to optimize what information to store and display in this page and allows the users to edit their own respective information. The profile and edit profile pages consist of displaying user information, their messages from other users, their site activities, and sign contracts that they are supposed to if they win or if their posts ended.

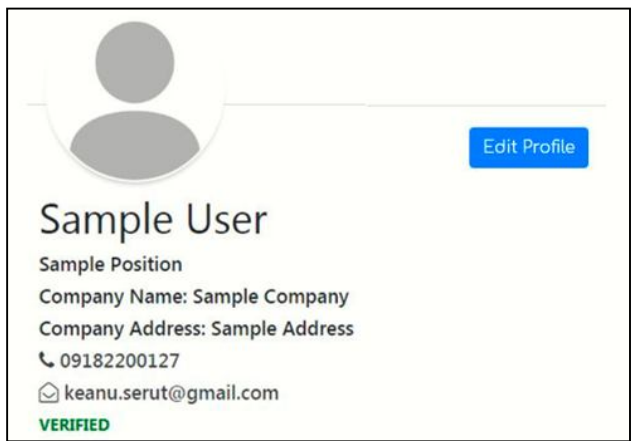

**Figure 8:** Procure Chain Profile page

**4. LIMITATIONS**

The complicated functions and scripts on the platform, specifically materialized the focus of the study to create an immutable system for purchasing goods and services. Therefore, the functionality of the system is limited to the basic processes in procurement, particularly in the declaration of winners, that will solely rely on the lowest bid price. Certain functions will sometimes fail when using internet browsers other than Google Chrome because of the web development languages used. The program relies heavily on the users' internet connection speeds, which may sometimes fail due to slow internet connections. Lastly, the procured items can only consist of items and fixed services since the input data cannot be changed for considerable specifications.

**5. DISCUSSION**

To further improve the system, the proponents allowed professionals in the field of procurement to test the system. The said individuals were then asked to give their opinion regarding the system and provide recommendation for the further improvement thereof. The recommendations were focused on the improvement of the interface to magnify the user-friendliness of the platform especially on the market page. The platform must also be compatible to any browser and device. Additionally, the automation of the Know Your Customer Verification (KYC) will increase the efficiency of the system. The proponents also notice a need to add variables for the declaration of procurement winners, not just the bid price. And lastly, the differentiation between individual users and private companies.

**6. CONCLUSION**

Different approach and techniques were used in different aspects and functions of the online procurement system. Different programming languages took part in building the system from the aesthetics up to the programming logic and blockchain integration.

The developed online procurement system provided a responsive, aesthetic site, functional scripts, and complex programming logic and design. The blockchain integration took place within the implementation of procurement processes, from posting of procured items or goods, up to the signing of contracts by the winner and the procurer. The account system and personalization of profiles can be seen from the users signing up and in editing of their profile. The system is fully-functional, maintained, and updated.
.

**ACKNOWLEDGEMENT**

The proponents of this study would like to thank the University Research and Development Services Office of the Technological University of the Philippines for the financial support.